\documentclass[aps,prd,10pt,nofootinbib,twocolumn,unsortedaddress]{revtex4}

\usepackage{epsfig}
\usepackage{bm}
\usepackage{latexsym}
\usepackage{natbib}
\usepackage{url}
\usepackage{dcolumn}
\usepackage{color}
\usepackage[usenames,dvipsnames]{xcolor}
\usepackage{amsfonts,amssymb,amsmath}
\usepackage{graphicx,epsfig}
\usepackage{psfrag}
\usepackage{subfigure}
\usepackage{comment}
\usepackage{hyperref}
\hypersetup{colorlinks=true}

\def\mpc{{\rm Mpc}}
\def\calT{{\cal T}}
\def\mathcalT{{\cal{T}}}
\def\Minf{M_{\rm inf}}
\def\Hinf{H_{\rm inf}}

\def\grel{g_{\rm rel}}
\def\be{\begin{equation}}
\def\ee{\end{equation}}
\def\gev{{\rm \,Ge\kern-0.125em V}}

\newcommand{\gsim}{\mbox{\raisebox{-.6ex}{~$\stackrel{>}{\sim}$~}}}

\begin{document}
 
\title{Constraints on just enough inflation preceded by a thermal era}

\author{Suratna Das}%
 \email{suratna@iitk.ac.in}
\affiliation{Indian Institute of Technology, Kanpur 208016, India}

\author{Gaurav Goswami}%
 \email{gaurav.goswami@ahduni.edu.in}
\affiliation{Institute of Engineering and Technology, Ahmedabad University, Navrangpura, Ahmedabad 380009, India}

\author{Jayanti Prasad}%
 \email{jayanti@iucaa.ernet.in}
\affiliation{Inter-University Centre for Astronomy and Astrophysics, Post Bag 4, Ganeshkhind, Pune 411007, India \\
Centre for Modeling and Simulation, Savitribai Phule Pune University, Pune 411007, India}

\author{Raghavan Rangarajan}%
 \email{raghavan@prl.res.in}
\affiliation{Theoretical Physics Division, Physical Research Laboratory, Navrangpura, Ahmedabad 380009, India}

\date{\today}

\begin{abstract}

If the inflationary era is preceded by a radiation dominated era in which the 
inflaton too was in thermal equilibrium at some very early time then the CMB data
places an upper bound on the comoving temperature of the (decoupled)
inflaton quanta.  
In addition, if one considers models of ``just enough" inflation,
where the number of e-foldings of inflation is just enough to solve the horizon
and flatness problems,
then we get 
a lower bound on the Hubble parameter during
inflation, $H_{\rm inf}$, which is in severe conflict with the upper bound from tensor
perturbations. 
Alternatively, imposing the upper bound on $H_{\rm inf}$ 
implies
that such scenarios are compatible with the data only if 
the number of relativistic degrees of freedom in the thermal bath in the pre-inflationary
Universe is extremely large (greater than $10^9$ or $10^{11})$.
We are not aware of scenarios in which this can be satisfied.

\end{abstract}

\pacs{}

\maketitle


 \section{Introduction}

The era in the early history of the Universe before cosmic inflation is completely unknown.
It has been argued \cite{Freivogel:2005vv,2013arXiv1309.4060B,2014JCAP...12..019B} 
that before the era of observable inflation, the Universe could have been sitting at a 
local minimum (metastable vacuum) in the effective potential of the underlying microscopic theory and 
then have tunnelled (via bubble nulceation) to a sufficiently flat direction causing slow-roll inflation.
The standard thermal history of the Universe begins after the end of inflation when the process 
usually referred to as reheating causes the Universe to get into a thermal state. 

In this work, we consider a different picture of the early Universe. 
In our picture, the pre-inflationary early Universe is in a thermal radiation dominated state but as the Universe expands, 
the density of radiation decreases and eventually falls below the energy 
density corresponding to the rolling field just like the energy density of the Universe 
gets dominated by a non-zero cosmological constant or quintessence field at late times.
We ask what can the present cosmic microwave background (CMB) data tell us about this possibility.

One crucial consequence of a pre-inflationary thermal history is that the  quanta of the
inflaton fluctuations $\delta \phi$ (the excitations about the classical inflaton background) 
could have been in thermal equilibrium very early in the pre-inflationary radiation era.
If so, it would then
subsequently decouple but would 
retain a thermal Bose-Einstein distribution even after decoupling. Hence the field $\delta \phi$ during
inflation
would be in a thermal state 
($\langle a_{\bf k}^\dagger a_{\bf k'}\rangle= [\exp(k/T)-1]^{-1}\delta^3({\bf k}-{\bf k'})$, where $T$ is the inflaton
comoving temperature),
rather than in the Bunch-Davies vacuum state.
This effect was studied in Ref.~\cite{2006PhRvL..96l1302B} where it was found that the
primordial scalar power spectrum gets an additional factor of $\coth [k/(2T)]$ and
the CMB temperature anisotropies get enhanced 
at large angular scales by an amount which depends upon the comoving
temperature of the inflaton 
quanta during inflation
\footnote{In Refs.~\cite{1993PhRvD..48..439G}, \cite{2006PhRvL..97y1301B} and \cite{Zhao:2009pt} a thermal state of 
decoupled gravitons was considered which enhances the temperature anisotropy and the B mode polarisation angular
power spectrum of the
CMB at large angles.}.
From the measurements of temperature anistropies of the CMB, one can put an upper limit on 
the comoving temperature of the inflaton.
This then has implications for the duration of inflation \cite{2006PhRvL..96l1302B}.
A more detailed analysis, 
using COSMOMC,
and including the modification of the inflaton mode functions due to a pre-inflationary
radiation era, has been carried out by us
in Ref.~\cite{Das:2014ffa}.  
However 
in Ref.~\cite{Das:2014ffa}
we 
treat both the comoving temperature $T$ and 
the number of e-foldings of inflation more than that required to solve
the horizon and flatness problems ($\delta N$)
as independent parameters.
We obtain a stronger constraint on the comoving temperature than in
Ref. \cite{2006PhRvL..96l1302B}. 
Furthermore we find that the best fit value of the number of e-foldings of inflation
is very close to the minimum duration of inflation 
required to solve
the horizon and flatness problem (though the improvement in likelihood is not very much
over the standard power law spectrum).

Other previous studies of a scenario
with a pre-inflationary radiation dominated era have also shown that the CMB data favours
scenarios with just enough inflation to solve the horizon and flatness problems (because of the low power
on large scales in the $TT$ angular power spectrum) \cite{2003JCAP...09..010C,
Powell:2006yg,2008JCAP...01..002N,Marozzi:2011da,
Cicoli:2014bja,
Hirai:2004kh,
Hirai:2005tn,Hirai:2007ne} (see however Ref. \cite {Wang:2007ws}). 
We point out
that there also exist  good theoretical reasons for considering models of ``just enough inflation" 
 in which $\delta N$ is small.  By considering anthropic bounds on the value of the curvature and
 including certain statistical arguments on the parameters of inflation, it is argued 
 in Ref. \cite{Freivogel:2005vv}
 that the 
 number of e-foldings of inflation should not be much larger than the minimum
 required by observations.  In Ref. \cite{Gibbons:2006pa} it is shown that 
 by considering a certain natural canonical measure on the space of all 
 classical universes the probability for $N$ e-foldings of inflation is
 suppressed by a factor of $\exp(-3N)$.
 Moreover, more inflation means that the inflaton potential is required to stay flat for a huge range of 
field values and can require super-Planckian excursions. 
Therefore below we shall a consider a scenario of just enough inflation.

None of the studies in Refs. \cite{2003JCAP...09..010C,
Powell:2006yg,2008JCAP...01..002N,Marozzi:2011da,
Cicoli:2014bja,
Hirai:2004kh,
Hirai:2005tn,Hirai:2007ne,Wang:2007ws}
consider the possibility that the inflaton too could have been
in thermal equilibrium during the pre-inflationary radiation era, as has been discussed above. 
In the scenario of Ref. \cite{Das:2014ffa} we
 presume that the inflaton  was in thermal equilibrium at some early time in the 
pre-inflationary radiation era 
and that it subsequently went out of equilibrium and decoupled, and so include a (frozen) thermal
Bose-Einstein distribution for the quanta of the inflaton field. 
We believe this is a natural extension of scenarios with a pre-inflationary radiation dominated era.  After all, the inflaton must have some
couplings to other fields 
-- some are needed 
for sufficient reheating to occur after inflation.

Our 
current
analysis below shows 
that such 
a scenario with just enough inflation and a frozen thermal distribution for the
inflaton quanta is
compatible with the CMB data only if the 
effective number of relativistic degrees of freedom for the pre-inflationary radiation plasma is extremely
large, in fact,  larger than $10^9$ or $10^{11}$, depending on whether or not any entropy is transferred to
the radiation after the inflaton decouples. 
This had not been recognised in Ref. \cite{Das:2014ffa}.

\section{The model}

Previous studies of pre-inflationary radiation era had either considered modifications of the mode functions of the inflationary scalar perturbations due to presence of the prior radiation era \cite{2003JCAP...09..010C,
Powell:2006yg,2008JCAP...01..002N,Marozzi:2011da,
Cicoli:2014bja,
Hirai:2002vm,Hirai:2003dh,
Hirai:2004kh,
Hirai:2005tn,Hirai:2007ne,Wang:2007ws}, or a thermal distribution of inflaton fluctuations 
associated with the inflaton being in thermal equilibrium at some very early phase in the radiation era
 \cite{2006PhRvL..96l1302B}. While the first contribution lowers the quadrupole moment of  the CMB $TT$ anisotropy spectrum, 
 the other contribution enhances the power at large angular scales. We, in Ref. \cite{Das:2014ffa}, included both these effects as a consistent approach to determining the signatures of a pre-inflationary radiation era. 
We obtained the scalar power spectrum in Ref. \cite{Das:2014ffa} as 
\begin{equation} 
\mathcal{P}_{\mathcal R}(k)=A\left(\frac {k}{k_P}\right)^{n_s-1}\coth\left(\frac{k}{2T}\right)
{\cal C}(k)\,,
\label{pi_pps1}
\end{equation}
where ${\cal C}(k)$ reflects the modification of the inflaton mode functions and hence the power spectrum due to a pre-inflationary
radiation era.  The modified mode functions are obtained by setting Bunch-Davies initial conditions on the inflaton field
in the pre-inflationary radiation dominated era, and matching solutions for the field fluctuations and their time derivatives 
at the (instantaneous) transition from 
the radiation to the inflationary era (as opposed to simply setting the Bunch-Davies initial conditions in the inflationary era). 
Denoting 
the physical temperature of the inflaton by 
${\cal T}(t)$, the comoving temperature $ T = a(t) {\cal T}(t)$, with $a=1$ today.
The $\coth\left(\frac{k}{2T}\right)$ factor appears due to the thermal distribution of the inflaton fluctuations:
$1+ 2 n_k=1+2[\exp(k/T)-1]^{-1}=\coth\left(\frac{k}{2T}\right)$.  
If one considers only the modification of the inflationary mode function due to the presence of a pre-inflationary radiation era 
\cite{Powell:2006yg, 2008JCAP...01..002N,Marozzi:2011da,
Cicoli:2014bja,
Hirai:2002vm,Hirai:2003dh,
Hirai:2004kh,
Hirai:2005tn,Hirai:2007ne,
Wang:2007ws}
then the power spectrum would not contain the $\coth[k/(2T)]$ factor, 
and similarly considering only the thermal distribution of the inflaton quanta  the power spectrum would not contain the ${\cal C}(k)$ factor \cite{2006PhRvL..96l1302B}.

We carried out a detailed COSMOMC analysis in Ref. \cite{Das:2014ffa}, using Planck 2013 and WMAP nine year data
and the power spectrum in Eq. (\ref{pi_pps1}).  In addition to the 6 standard parameters
for the Monte Carlo simulation
we included
two extra parameters -- $\delta N$, the number of e-foldings of inflation in excess of the minimum required to solve the horizon
and flatness problems, and $T$, the comoving temperature.
(The factor ${\cal C}(k)$ in the power spectrum 
depends on $\delta N$ through
the time when the change from the radiation era to the inflationary era takes place.)
For an inflationary scale of $10^{15}\gev$ we found that the best fit value of $\delta N$ is 0.08 (the marginalized value is consistent with zero with 68 \% confidence).  This indicates that the data favours a `just enough' inflationary scenario.
As discussed earlier, there are also theoretical motivations for considering such a 
scenario.
We also obtained an upper limit on the comoving temperature as $T< 1.3\times10^{-4}\,\mpc^{-1}$ at 68\% C.L. (for the GUT and the electroweak scale).

We will now show that 
this scenario with a frozen
thermal distribution of inflaton fluctuations and `just enough' inflation preceded by a radiation dominated era faces a severe
conflict as it either implies a very large Hubble parameter during inflation that conflicts with
the known upper bound,  or requires an extremely large  
number of relativistic degrees of freedom (greater than $10^9$ or $10^{11})$ in the early Universe.


\section{Analysis}

In the following analysis 
we use the subscript 0 to refer to  
the present epoch.
Let $t_*$ be the epoch when the mode $k_0$ corresponding to our current Hubble radius was crossing
the Hubble radius during inflation.  Then 
 \begin{equation} \label{eq:k*}
  k_0 = a_* H_* = a_0 H_0 \; ,
 \end{equation}
 where $a_* = a(t_*)$, $H_* = H(t_*)$, $H(t_0)=H_0$ and $a_0=1$.
 Presuming that the Hubble parameter is approximately constant during inflation, we set
 $H_* = H_{\rm inf}=M_{\rm inf}^2/(\sqrt3M_{\rm P})$, where $M_{\rm inf}$ and 
 $M_{\rm P}=2.4\times10^{18}\gev$ are the energy
 scale of inflation and the reduced Planck mass respectively.
 Then
 \begin{eqnarray}
 \calT_*&=&T/a_*\nonumber\\
 &=&\frac{T H_{\rm inf}}{H_0}\,.
 \label{Tstar}
 \end{eqnarray}
 If there were $\delta N$ e-foldings of inflation before $t_*$ then
 \be
 \calT_i=\calT_* \exp\delta N\,,
 \label{Ti}
 \ee
 where $\calT_i$ is the physical inflaton temperature at the beginning of inflation.

We assume that the radiation energy density 
at the onset of inflation at $t_i$
is equal to that of the inflaton, i.e.
$\rho_r = \rho_\phi$.  
Then the radiation temperature at $t_i$
    \begin{eqnarray}\label{eq:V}
  \calT_\gamma&=& \left( \frac{30}{\pi^2 g_{\rm rel}} \right)^{1/4} {M_{\rm inf}}
  \label{TMinf}\\
  &=&\left( \frac{90}{\pi^2 g_{\rm rel}} \right)^{1/4}
  \sqrt{\Hinf M_{\rm P}}\,.
  \label{THinf}
   \end{eqnarray}
 Here $g_{\rm rel}$ is the effective number of relativistic degrees of freedom
 for the plasma filling the Universe at 
 the onset of inflation. 
 If the underlying microscopic theory is the Standard Model of elementary particle physics, 
 $g_{\rm rel} = 106.75$ while if it is MSSM, then $g_{\rm rel} = 228.75$
 \cite{1990eaun.book.....K}.
 If the number of relativistic degrees of freedom 
 stays the same between the epoch of decoupling of the inflaton and the epoch of 
 onset of inflation (i.e. no species annihilates or becomes non-relativistic in between),
 we can assume that the inflaton temperature equals the radiation temperature at the onset of inflation,  i.e. 
 \be
 {\cal T}_{i} = {\cal T}_\gamma\,.
 \label{Tigamma}
 \ee 
 Then
 combining Eqs. \eqref{Tstar}, \eqref{Ti}, \eqref{THinf} and \eqref{Tigamma},
 we get
 \begin{equation} \label{Hinf:general}
 \frac{H_{\rm inf}}{M_{\rm P}} =  
 \left( \frac{90}{\pi^2 g_{\rm rel}} \right)^{1/2} 
 \left( \frac{H_0}{T} \right)^2 
 e^{-2 \delta N} \; .
\end{equation}

Below we shall take
\be
T< 1\times10^{-4}\mpc^{-1}\,.
\label{Tboundnew}
\ee
As mentioned above for GUT scale inflation the best fit value of 
 $\delta N$ is $0.08$.  Assuming the energy scale to be much smaller (e.g. the electroweak scale) implies 
 that the best fit $\delta N$ is $0.02$.
 If we assume that $\delta N \approx 0.05$ 
 and $T < 10^{-4} {\rm Mpc}^{-1}$,
 and take $H_0\approx({4400} \,\mpc)^{-1}$ 
 \cite{Ade:2013zuv}
  and
 $g_{\rm rel} \sim100$, then Eq. (\ref{Hinf:general}) implies that
\begin{equation} 
  \label{Hinf:prediction}
  \frac{H_{\rm inf}}{M_{\rm P}} >{1} \; ,
\end{equation}
or,
\be
   \Minf> M_{\rm P} \, .
\ee
But from the upper bound on $r$ at $k= 0.002\, \mpc^{-1}$ (v.i.z. $r < 0.1$), we have
\cite{Ade:2013uln} 
\begin{equation} \label{Hinf:observation}
 \frac{H_{\rm inf}}{M_{\rm P}} < 
 3.7\times 10^{-5}
 \; ,
\end{equation}
or $\Minf<1.9\times10^{16}\gev$.
Clearly the lower bound on $\Hinf$ in Eq.~\eqref{Hinf:prediction} is in severe conflict with the
CMB constraint. (Note that for $k = 0.002 \,\mpc^{-1}$ the coth factor and $\cal C$ in 
Eq.~(\ref{pi_pps1}) reduce to $\sim1$ and hence we can use Eq. \eqref{Hinf:observation}.)

 As is clear from Eq.~(\ref{Hinf:general}) this result is a consequence of picking a small value of 
 $\delta N$ (larger values of $\delta N\gsim 6$ will decrease the lower bound on $\Hinf$ in
 Eq.~\eqref{Hinf:prediction} to acceptable levels).
 Our result is thus a serious constraint on inflationary models with just enough inflation which are preceded with a 
 pre-inflationary radiation era where one assumes that the inflaton too could have been in thermal equilibrium at some
 early time prior to inflation. 

We now study possible solutions to this conundrum. 
Eq. \eqref{Hinf:general} implies that
Eq. (\ref{Hinf:prediction}) can be recast as
\begin{equation} 
  \label{Hinf:prediction2}
 \frac{H_{\rm inf}}{M_{\rm P}}  > \frac{10}{\sqrt {g_{\rm rel}}}  \; .
\end{equation}
 However, Eq. \eqref{Hinf:observation} then implies that 
 \begin{equation}
 \grel>10^{11}\,.
 \end{equation}
 This is an extremely
 large number of relativistic degrees of freedom in the early Universe.

At this point it is worthwhile recalling another assumption we have made in arriving at our result.
As Eq. \eqref{Tigamma} indicates, we have assumed that the temperature of 
the decoupled inflaton quanta and that of the pre-inflationary radiation plasma is the same.
This assumption will break down if between the epoch of decoupling of the inflaton (at time $t_{\rm dec}$) 
and the onset of inflation (at time $t_i$) some species annihilates and hence causes the temperature of the radiation 
to be more than that of the inflaton at $t_i$. Let us see what are the consequences of 
relaxing the assumption of Eq. \eqref{Tigamma}.

We now suppose that at the Planck time, the inflaton, BGUT particles, and GUT particles were in thermal equilibrium
(BGUT refers to Beyond GUT), possibly via gravitational interactions.  
After the Planck scale, at time $t_{\rm dec}$, the inflaton decouples.  A bit later, at a time $t_1$, 
the BGUT particles become non-relativistic  and annihilate 
in equilibrium 
and transfer their entropy to the GUT particles but not to the inflaton (since it is already decoupled).
Using $b$ and $a$ for before and after (instantaneous) annihilation, conservation of entropy implies that
\begin{equation} \label{eq:entr_cons}
\frac{\mathcalT_b}{\mathcalT_a}=\left(\frac{g_{*a}}{g_{*b}}\right)^\frac13 \; ,
\end{equation}
where $g_{*b}$ is the number of relativistic degrees of freedom before the annihilation and 
$g_{*a}$ is the number of relativistic degrees after the annihilation.

Assuming no further significant entropy production from $t_1$ till $t_i$,
the temperature of the 
inflaton quanta
at the onset of inflation
is
\begin{equation}
\mathcalT_i = \mathcalT_b (a_{1}/a_i) \; ,
\end{equation}
while the temperature of the radiation 
evolves to
\begin{eqnarray}
\mathcalT_{\gamma} &= &\mathcalT_a (a_{1}/a_i) \\
 \label{Tgamma1}
&=& \mathcalT_b \left(\frac{g_{*b}}{g_{*a}}\right)^\frac13 (a_{1}/a_i)
\end{eqnarray}
and
$g_{\rm rel}=g_{*a}$.
Then 
\begin{equation}
\mathcalT_i=(\mathcalT_b/\mathcalT_a) \mathcalT_{\gamma} \; ,
\end{equation}
and using Eq. \eqref{eq:entr_cons} and the fact that $\rho_\phi = \rho_\gamma$ at the beginning of inflation we get
\begin{equation}
\mathcalT_i =\left(\frac{g_{*a}}{g_{*b}}\right)^\frac{1}{3}
\left(\frac{90}{\pi^2 g_{*a}}\right)^\frac{1}{4} \,
\sqrt{\Hinf M_{\rm P}}\,.
\label{newTi}
\end{equation}
Combining Eq. (\ref{newTi}) 
with Eqs. \eqref{Tstar} and \eqref{Ti},
and using the bounds in Eqs. \eqref{Tboundnew} and \eqref{Hinf:observation},
and setting $\delta N=0.05$ and $H_0\approx ({4400} \,\mpc)^{-1}$
gives 
\be
g_{*b}>{2}\times10^8 g_{*a}^{1/4} \approx 10^9\,
.
\label{gbbound}
\ee
Such a large change in $g_*$ can not be obtained from the annihilation of a single species.
However if multiple species 
annihilate
in equilibrium at
$t_1, t_2, .., t_n$ and transfer their entropy to radiation, then
\begin{eqnarray}
\mathcalT_i&=&\mathcalT_{b1} (a_{1}/a_i),\\
\mathcalT_{\gamma} &=&
\mathcalT_{a1}\left(\frac{\tau_{a2}}{\tau_{b2}}\right)
..\left(\frac{\tau_{an}}{\tau_{bn}}\right)
\left(\frac{a_1}{a_2}\right)
\left(\frac{a_2}{a_3}\right)..
\left(\frac{a_n}{a_i}\right)\nonumber\\
&=&\mathcalT_{b1}
\left(\frac{g_{*b1}}{g_{*a1}}\right)^\frac13
\left(\frac{g_{*b2}}{g_{*a2}}\right)^\frac13
..\left(\frac{g_{*bn}}{g_{*an}}\right)^\frac13
\left(\frac{a_1}{a_i}\right)\nonumber\\
&=&\mathcalT_{b1}\left(\frac{g_{*b1}}{g_{*an}}\right)^\frac13(a_1/a_i)\,,
\end{eqnarray}
as in Eq. \eqref{Tgamma1} with $g_{*an}=g_{\rm rel}$.
$(g_{*b\,{m+1}}=g_{*am}.)$
Then we obtain the same result as in Eq. (\ref{gbbound}) for $g_{*b1}$, i.e. 
the number of relativistic degrees of freedom before the annihilations must be greater than $10^9$.

We thus conclude 
that any scenario of just enough inflation, in which 
the pre-inflationary universe was radiation dominated and in which the inflaton quanta 
too were in thermal 
equilibrium at some early epoch, requires an extremely large number of relativistic species to be present in the pre-inflationary radiation plasma to be consistent with the CMB data.

 \section{Discussion and Conclusion}  
 
 We now present a physical interpretation of our results.  The $\coth[k/(2T)]$ factor in our power
 spectrum leads to an enhancement of power on large scales.  To suppress this requires putting
 an upper bound on the comoving temperature $T$.  
 If we express the coth factor 
 for the current horizon scale as $\coth[k_0/(2a_*\calT_*)]=\coth[H_{\rm inf}/(2\calT_*)]$, then 
 the upper bound on $T$ translates into
 an upper bound 
 on $\calT_*$.  For a small value of $\delta N$, i.e., if the duration 
 of inflation is just enough to solve the horizon and flatness problems, 
 $\calT_i\approx \calT_*$.  Now if the (frozen) inflaton temperature 
 at the beginning of inflation is the same
 as that of the thermal radiation in the pre-inflationary radiation era, then one gets an upper bound
 on the radiation temperature 
 \begin{equation}
 \calT_\gamma<4\times10^{13}\gev \frac{H_{\rm inf}}{9\times10^{13}\gev}\,.
 \end{equation}
 To satisfy this low bound on the pre-inflationary radiation temperature and the condition 
 $\rho_{r}=\rho_\phi=3H_{\rm inf}^2 M_{\rm P}^2$ at the beginning of inflation then requires $g_{\rm rel}$ to be larger than  
$10^{11}$.
 Alternatively, the radiation temperature at the beginning of inflation can be much larger
 than the inflaton temperature if entropy was released in the Universe after the
 inflaton decoupled.  Then while $\calT_i$ will be less than $4\times 10^{13}\gev$,
 $T_\gamma$ can be as high as
 $1\times10^{16}\gev$ with $g_{\rm rel}\sim100$.  But then $g_*$
 before entropy release is $(\calT_\gamma/\calT_i)^3 g_{\rm rel}\approx 10^9$.

In models of thermal inflation invoked to solve the moduli problem \cite{Lyth:1995hj, Lyth:1995ka, Randall:1994fr} primordial inflation is followed by a later phase of `thermal' inflation which is driven by the potential energy of a `flaton' field trapped at the origin at the minimum of its thermal effective potential \cite{Lyth:1995hj, Lyth:1995ka,Hiramatsu:2014uta}.  Density perturbations generated during the short later phase of  thermal inflation influence the spectrum only on very small scales while the larger scale perturbations which influence the 
large scale structure and cosmic microwave background are largely generated during the earlier primordial inflation \cite{Hong:2015oqa}.  Since the flaton has thermal interactions one would expect that the flaton quanta would have a thermal Bose-Einstein distribution, which has so far not been considered.    
In models of warm inflation a thermal bath is generated by dissipation of the inflaton field and is maintained in the Universe during inflation.
If the inflaton fluctuations are also in thermal equilibrium with the thermal bath there will again be a coth term in the density perturbation spectrum, as seen, for example, in Refs. \cite{Ramos:2013nsa,Bartrum:2013fia,Bastero-Gil:2014jsa}.  In both thermal and warm inflation, the presence of the coth term would affect  density perturbations with large power on the relevant large scales, as in 
Ref. \cite{2006PhRvL..96l1302B}, and can also lead to further constraints as discussed above if one also imposes a finite duration of inflation.  We plan to study these scenarios in greater detail in the future.

In conclusion, we have explored a scenario of inflation preceded by a radiation dominated era.
We presume that the inflaton too could have been in thermal equilibrium with the radiation at
some very early era and so we consider a frozen 
thermal state for the inflaton field with a Bose-Einstein
distribution for the (decoupled) inflaton quanta.  There is then an upper bound on the comoving
relic temperature for the inflaton \cite{2006PhRvL..96l1302B,Das:2014ffa}. 
Then, in these inflation scenarios and with just enough e-foldings
to solve the horizon and flatness problem,
which is motivated by constraints on large inflaton
field excursions  
and are also indicated by the CMB data, 
we find  
a lower bound on
the Hubble parameter during inflation which is in severe conflict with the upper bound on
the Hubble parameter from CMB observations.  
To resolve this conflict
one has to 
allow for an extremely large 
number of relativistic degrees of freedom in the thermal bath in the very early Universe,
greater than $10^9$ or $10^{11}$ depending on whether or not there has been entropy transfer to radiation
after the inflaton decoupled.
We are not aware of any scenario that may give such large values of $g_*$ in the early Universe.

\begin{acknowledgements}
R.R. would like to thank S. Minwalla and S. Trivedi for helpful discussions.
We would like to thank the 
 first referee for constructive comments about our work that helped to improve the presentation of our results.
 We would also like to thank the second referee for the suggestion to apply 
 our analysis to models of thermal and warm inflation.
 Numerical work for the present study was done on the IUCAA HPC facility. 
 Work of S.D. is supported by Department of Science and Technology, 
 Government of India under the Grant Agreement number IFA13-PH-77 (INSPIRE Faculty Award). 
 J.P. would like to thank the Science and Engineering Research Board (SERB) of the 
 Govt. of India for financial support via a Start-Up Research Grant (Young Scientists)
 SR/FTP/PS-102/2012. 
 \end{acknowledgements}

\label{Bibliography}
 \bibliographystyle{h-physrev3}
 \bibliography{Hinf}

\end{document}